\documentclass[a4paper]{jpconf}
\usepackage{graphicx}
\usepackage{xspace}

\begin{document}
\title{APEIRON: composing smart TDAQ systems for high energy physics experiments}

\author {
Roberto~Ammendola$^2$,
Andrea~Biagioni$^1$,
Carlotta~Chiarini$^{3,1}$,
Andrea~Ciardiello$^{3,1}$,
Paolo~Cretaro$^1$,
Ottorino~Frezza$^1$,
Francesca~Lo~Cicero$^1$,
Alessandro~Lonardo$^1$,
Michele~Martinelli$^1$,
Pier~Stanislao~Paolucci$^1$,
Cristian~Rossi$^1$,
Francesco~Simula$^1$,
Matteo~Turisini$^1$,
Piero~Vicini$^1$
}

\address{$^1$ Istituto Nazionale di Fisica Nucleare (INFN), sezione di Roma, Rome, Italy}
\address {$^2$ Istituto Nazionale di Fisica Nucleare (INFN), sezione di Roma Tor Vergata, Rome, Italy}
\address {$^3$ Dipartimento di Fisica, Sapienza Università di Roma, Rome, Italy}

\ead{alessandro.lonardo@roma1.infn.it}

\begin{abstract}
APEIRON is a framework encompassing the general architecture of a distributed heterogeneous processing platform and the corresponding software stack, from the low level device drivers up to the high level programming model.
The framework is designed to be efficiently used for studying, prototyping and deploying smart trigger and data acquisition (TDAQ) systems for high energy physics experiments. 
\end{abstract}

\section{Introduction}
The general architecture of the APEIRON distributed processing platform includes \textit{m} data sources, corresponding to the detectors or sub-detectors, feeding a sequence of \textit{n} stream processing layers, making up the whole data path from readout to trigger processor (or storage server).

\begin{figure}
\begin{center}
\includegraphics[width=0.6\textwidth]{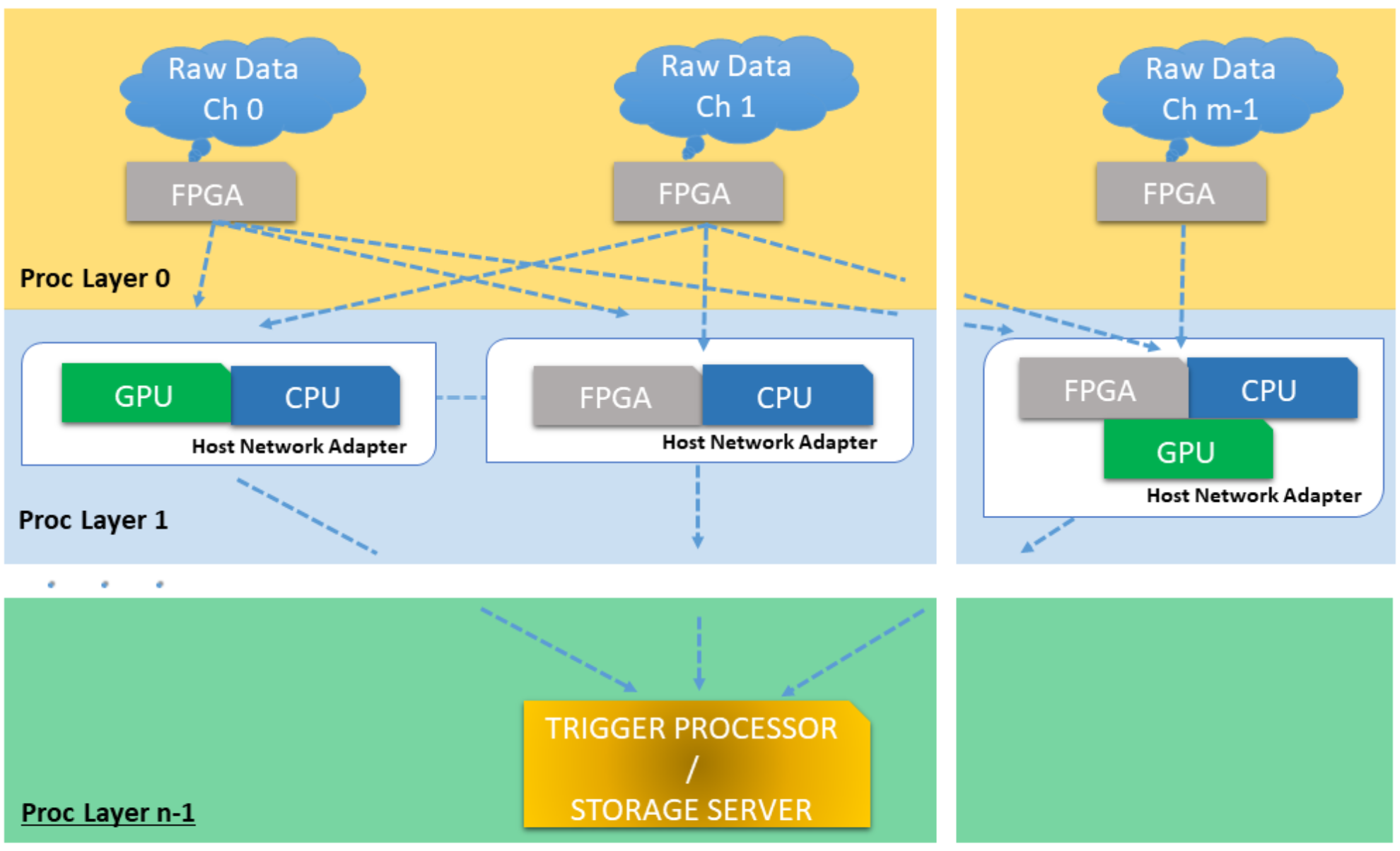}
\end{center}
\caption{\label{proc_layers}Recombination of data streams through processing layers.}
\end{figure}

The processing platform features a modular and scalable low-latency network infrastructure with configurable topology. This network system represents the key element of the architecture, enabling the low-latency recombination of the data streams arriving from the different input channels through the various processing layers, as shown in Figure~\ref{proc_layers}.

Developers can define scalable applications using a dataflow programming model (inspired by Kahn Process Networks~\cite{gilles1974semantics}) that can be efficiently deployed on a multi-FPGAs system: the APEIRON communication IPs allow low-latency communication between processing tasks deployed on FPGAs, even if hosted on different computing nodes.

Thanks to the use of High Level Synthesis tools in the workflow, tasks are described in high level language (C/C++) while communication between tasks is expressed through a lightweight API based on non-blocking \textit{send()} and blocking \textit{receive()} operations.

\begin{figure}[hbt!]
\centering
  \begin{minipage}[t]{.65\textwidth}
\centering
    \includegraphics[width=.95\textwidth]{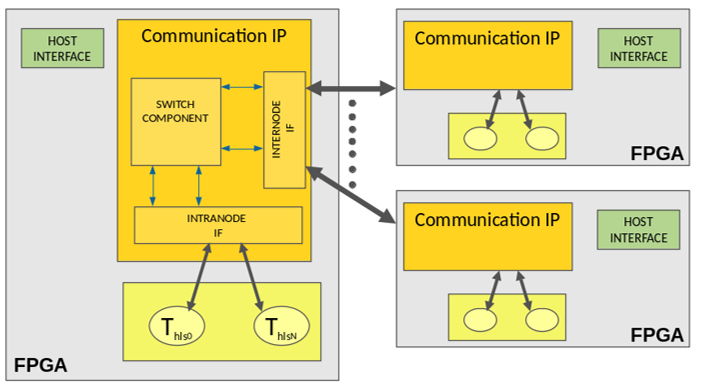}
    \caption{Communication IPs managing data streams I/O and communication between HLS computing tasks (represented as yellow ovals).}
    \label{fig:comm_ip}
  \end{minipage}
  \quad
  \begin{minipage}[t]{.25\textwidth}
\centering
    \includegraphics[width=.95\textwidth]{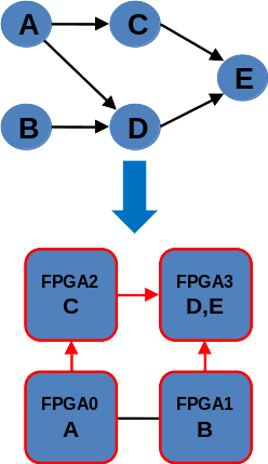}
    \caption{Deployment of five tasks on four interconnected FPGAs.}.
    \label{fig:mapping}
  \end{minipage}
\end{figure}

The mapping between the computational data flow graph and the underlying network of FPGAs, such as that shown in Figure~\ref{fig:mapping}, is defined by the designer with a configuration tool, by which the framework will produce all project files required for the FPGAs bitstream generation. The interconnection logic is therefore automatically built according to the application needs (in terms of input/output data channels) as shown in Figure~\ref{fig:comm_ip}, allowing the designer to focus on the processing tasks expressed in C/C++ .


The aim of the APEIRON project is to develop a flexible framework that could be adopted in the design and implementation of both "traditional" low level trigger systems and of data reduction stages in trigger-less or streaming readout experimental setups characterized by high event rates.
For this purpose we studied and implemented algorithms capable of boosting the efficiency of these classes of online systems based on Neural Networks (NN), trained offline on Tensorflow/Keras and leveraging the QKeras and HLS4ML~\cite{Duarte_2018} software packages for deployment on FPGA.
We have validated the framework on the physics use case represented by the partial particle identification system for the low-level trigger of the NA62 experiment~\cite{L0Paper}, working on data from its Ring Imaging Cherenkov detector to pick out electrons and number of charged particles. 

\section{Motivation}
As for the requirements imposed by applications in the class of real-time dataflow processing, FPGA devices are a good fit inasmuch as they can provide not only adequate computing, memory and I/O resources but also a smooth programming experience. High-Level Synthesis tools, after several years since their appearance, are quickly reaching a technological readiness that paves the way to the adoption of these reconfigurable accelerators by a class of users much broader to that composed by skilled developers used to employ Hardware Description Language-based workflows. 

The main motivation for the design and development of the APEIRON framework is that the currently available HLS tools do not natively support the deployment of applications over multiple FPGA devices, which severely chokes the scalability of problems that this approach could tackle. To overcome this limitation, we envisioned APEIRON as an extension of the Xilinx Vitis HLS framework able to support a network of FPGA devices interconnected with a low-latency direct network as the reference execution platform. 

\section{The APEIRON Framework}
The Communication IP is the evolution of the APEnet~\cite{APEnetTwepp:2013} and exanet~\cite{DSD:EXANEST:2017} designs for HPC systems and represents the main enabling component for the APEIRON framework, defined as the general architecture of an FPGA-based distributed stream processing platform and the corresponding software stack. 

\begin{figure}
\begin{center}
\includegraphics[width=0.7\textwidth]{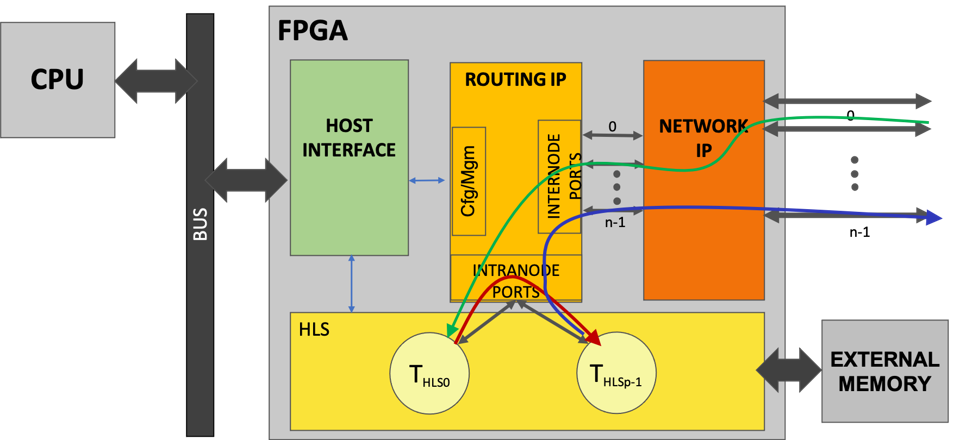}
\end{center}
\caption{\label{node} HLS kernels performing intra-node (red line) and inter-node (green line – receive, blue line– send) communications .}
\end{figure}

The Communication IP allows data transfers between processing tasks hosted in the same node (intra-node communications) or in different nodes (inter-node communications), see Figure~\ref{node}. In the context of the APEIRON framework, processing tasks are implemented by HLS kernels with Xilinx Vitis. The details of the interface between HLS kernels – the endpoints of the communication – and the Communication IP are described at the end of this section.
The Routing IP defines the switching technique and routing algorithm; its main components are the Switch component, the Configuration/Status Registers and the InterNode and IntraNode interfaces. 

The Switch component dynamically interconnects all ports of the IP, implementing a channel between source and destination ports. 

Dynamic links are managed by routing logic together with arbitration logic: the Router configures the proper path across the switch while the Arbiter is in charge of solving contentions between packets requiring the same port. 

For inter-node communications, the routing policy applied is the dimension-order one: it consists in reducing the offset along one dimension to zero before considering the offset in the next dimension. 

The employed switching technique — i.e., when and how messages are transferred — is Virtual Cut-Through (VCT)~\cite{Kermani79virtualcut-through:}: the router starts forwarding the packet as soon as the algorithm has picked a direction and the buffer used to store the packet has enough space. The deadlock-avoidance of DOR routing is guaranteed by the implementation of two virtual channels for each physical channel (with no fault-tolerance guaranteed)~\cite{Duato:1995:Deadlock}. 

The transmission is packet-based, meaning that the Communication IP sends, receives and routes packets with a header, a variable size payload and a footer. 

The Communication IP was co-designed with the APEIRON software stack in order to achieve very low-latency and scalable bandwidth (via IP design reconfiguration) between processing tasks defined as High-Level Synthesis Kernels. 

Starting from a YAML configuration file describing the attributes of each HLS kernel, namely its number of input and output channels and the IntraNode port of the Communication IP to which it is connected, the APEIRON framework links the Communication IP and the HLS kernels that are connected to it and generates the bitstream for the overall design. 

The only requisite that HLS kernels must satisfy is in the format of their prototype that must be in this form: 

\begin{verbatim}
void example_apeiron_task(
    [optional kernel-specific list of parameters] 
    message_stream_t message_data_in[N_INPUT_CHANNELS],
    message_stream_t message_data_out[N_OUTPUT_CHANNELS])
\end{verbatim}

In this way, the HLS kernel implements a generic stream interface for each communication channel, based on the AXI4-Stream protocol. The communication between kernels is expressed through a lightweight C++ API based on non-blocking send() and blocking receive() operations. This simple API allows the HLS developer to perform communications between kernels, either deployed on the same FPGA (intra-node communication) or on different FPGAs (inter-node communication) without knowing the details of the underlying packet communication protocol. 

\begin{figure}
\begin{center}
\includegraphics[width=0.7\textwidth]{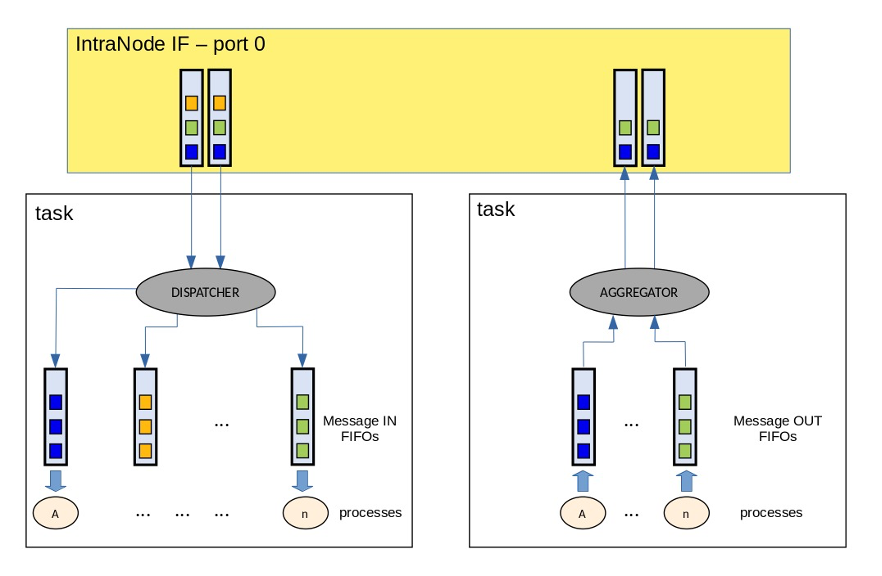}
\end{center}
\caption{\label{disp_aggr}Interface between Intranode Port 0 and the corresponding HLS Task (task\_id 0), Message IN FIFOs are identified by the ch\_id API parameter.}
\end{figure}

The Communication API can be represented with the following pseudo-code: 
\begin{verbatim}
    size_t send (msg, size, dest_node, task_id, ch_id);  
    size_t receive (ch_id); 
\end{verbatim}
where: 
\begin{verbatim}
dest_node is the n-Dim coordinate of the destination node (FPGA) in an n-Dim torus network; 
task_id is the local-to-node receiving task (kernel) identifier (0-3); 
ch_id is the local-to-task receiving FIFO (channel) identifier (0-127).
\end{verbatim}

The Communication Library leverages AXI4-Stream Side-Channels to encode all the information needed to forge the packet header. 

Adaptation toward/from IntraNode ports of the Routing IP is done by two APEIRON IPs: Aggregator and Dispatcher, shown in Figure~\ref{disp_aggr}. The Dispatcher receives incoming packets from the Routing IP and forwards them to the right input channel, according to the relevant fields of the header. The Aggregator receives outgoing packets from the task and forges the packet header, filling then the header/data FIFOs of the Routing IP. 

\section{Physics Use Case} 

\begin{figure}[hbt!]
\centering
  \begin{minipage}[t]{.45 \textwidth}
\centering
    \includegraphics[width=1.3\textwidth]{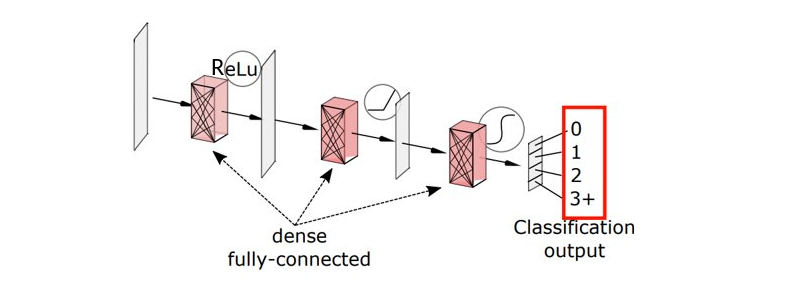}
    \caption{Dense Model (64x16x4) schematics.}
    \label{DenseModel}
  \end{minipage}
  \quad
  \begin{minipage}[t]{.5\textwidth}
\centering
    \includegraphics[width=0.9\textwidth]{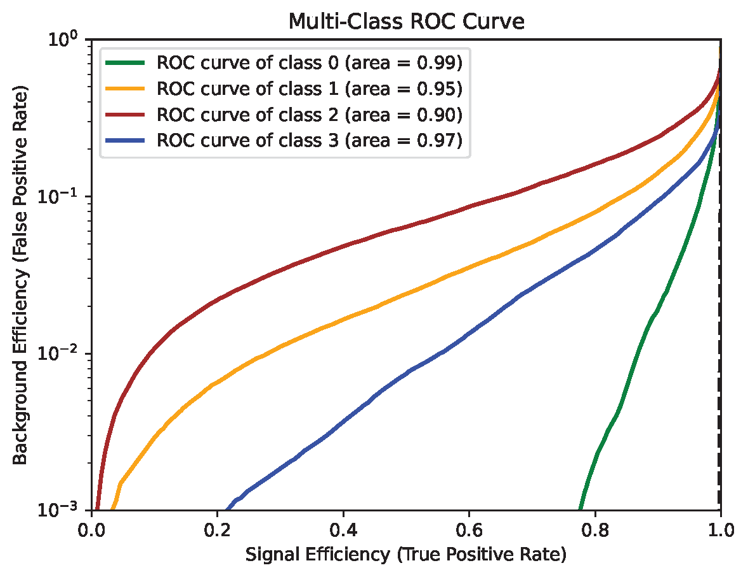}
    \caption{ROC curves for $N_r$}.
    \label{ROC}
  \end{minipage}
\end{figure}

NA62 is a fixed-target experiment at the CERN SPS North Area, dedicated to
measurement of rare kaon decays. We have designed FPGA-RICH, a Particle Identification (PID) system, based on the APEIRON framework and implemented on a single FPGA device, capable of providing results to the online trigger.\\
This systems represents the evolution of the GPURICH one that provided the same capabilities but on a more complex architecture, with a GPU performing a geometry based PID algorithm and a FPGA hosting the the NaNet design~\cite{NanetTwepp:2013, NaNet:TWEPP2014:paper,NanetTwepp:2015} implementing the low-latency direct data transfer between the detector and the GPU memory.\\
FPGA-RICH receives RICH detector events in a streaming fashion and performs the PID task using a neural network (NN), supporting a throughput greater than 10~MHz as per experiment specifications.
According to the APEIRON workflow the NN is implemented as a HLS Kernel and receives input data from the RICH detector only (\textit{seedless} model). The resulting model, depicted in Fig.~\ref{DenseModel}, is a three layer Dense network (64x16x4) having in input up to 64 normalized IDs of the PMTs hit by the Cherenkov photons in a single event. To limit the FPGA resources footprint we performed a quantization step on the model using QKeras, resulting in  two different fixed point representations:  $<$8,~1$>$ for weights and biases and $<$16,~6$>$ for activations.\\
Two different features can be inferred for each event: the number of charged particles ($N_r$) and the number of $e^\pm$ ($N_e$). In order to prepare the training and validation data for the NN, we prepared different data sets composed by events extracted from NA62 physics runs using the experiment analysis framework. The ground truth for training was provided by the seedless \textit{RichReco} offline reconstruction method.\\
Since the NN result would be used to enforce a trigger decision the inference performance of the NN is of utmost importance: to get a training set as much as possible similar to online data we trained the network with 3~Mevents extracted from run 8011. Validation has been done on 3.5~Mevents from run 8893 with satisfying results, as shown by ROC curves for $N_r$ in Fig.~\ref{ROC}. \\
Since the NA62 RICH detector is able to discriminate the kind of charged particles only in the $15-35$~$GeV/c$ energy range, results for $N_e$ are not equally satisfying.\\
The model was synthesized  on a Xilinx VCU118 FPGA platform at a 150~MHz clock frequency, and used a very limited amount of resources (14$\%$ LUT, 2$\%$ DSP), being able to sustain a 18.75~MHz throughput with a latency of 146.66~ns.

 
 
\section{Conclusions and Future Work} 
We are continuing the development of the APEIRON framework in order improve its performance and usability. We are finalizing the development of the FPGA-RICH system, integrating the NN kernel in the framework encouraged by the good performance on the identification of charged particles. We envisioned a solution to improve results in identification of  $e^\pm$, using the LKr calorimeter online primitives that provide information related to the energy of the event.


\section*{References} 
\bibliography{bibliography}

\end{document}